# Optical Response of Monolayer CdTe/CdS Quantum Dots to X-rays and Gamma-rays

*Girija Gaur [†], Dmitry S. Koktysh [‡, δ], Daniel M. Fleetwood [†], Robert A. Weller [†], Robert A. Reed [†], Bridget R. Rogers [ₑ] and Sharon M. Weiss [\*, †]*

[†] Department of Electrical Engineering and Computer Science, Vanderbilt University, Nashville, TN 37212, USA;

[‡] Vanderbilt Institute of Nanoscale Science and Engineering, Vanderbilt University, Nashville, TN 37212, USA

[δ] Department of Chemistry, Vanderbilt University, Nashville, TN 37212, USA

[ₑ] Department of Chemical and Biomolecular Engineering, Vanderbilt University, Nashville, TN 37212, USA;

ABSTRACT

We investigate the influence of X-ray and gamma-ray irradiation on the photophysical properties of sub-monolayer CdTe/CdS quantum dots (QDs) immobilized in porous silica ($PSiO_2$) scaffolds. The highly luminescent QD-$PSiO_2$ thin films allow for straightforward monitoring of the optical properties of the QDs through continuous wave and time-resolved photoluminescence spectroscopy. The $PSiO_2$ host matrix itself does not modify the QD properties. X-ray irradiation of the QD-$PSiO_2$ films in air leads to an exponential decrease in QD emission intensity, an exponential blue-shift in peak emission energy, and substantially faster exciton decay rates with increasing exposure doses from 2.2 $Mrad(SiO_2)$ to 6.6 $Mrad(SiO_2)$. Gamma-ray irradiation of a QD-$PSiO_2$ thin film at a total exposure dose of

700 krad($SiO_2$) in a nitrogen environment results in over 80% QD photodarkening but no concurrent blue-shift in peak emission energy due to a lack of photo-oxidative effects. Near-complete and partial reversal of irradiation-induced photodarkening was demonstrated on X-ray and gamma-ray irradiated samples, respectively, through the use of a surface re-passivating solution, suggesting that there are different contributing mechanisms responsible for photodarkening under different irradiation energies. This work contributes to improving the reliability and robustness of QD based heterogeneous devices that are exposed to high risk, high energy environments with the possibility of also developing QD-based large area, low-cost, re-useable, and flexible optical dosimeters.

## 1.   Introduction

Over the past decade, there have been tremendous improvements in the synthesis of colloidal quantum dots (QDs), enabling the artificial engineering of wavelength specific, strong light emitters with controllable surface chemistries. Due to their unique optical properties, QDs have found many applications ranging from photovoltaics,[1-5] photodetectors,[6-7] fluorescence probes,[8-10] and LEDs[11-15] to radiation scintillators[16-17], radiation oncology[18], and X-ray imaging screens.[19] For some of these applications, knowledge of the influence of highly energetic photons on the optical properties of QDs immobilized on a substrate is essential for achieving reliable and robust QD-based device operation over extended periods of time in high risk and high radiation environments. Several studies have examined the radiation hardness of QD-based devices grown by metal organic chemical vapor deposition or molecular beam expitaxy.[20-21] Very few studies have investigated the effects of high energy radiation on colloidal QDs. One study reported rapid degradation of CdSe/ZnS colloidal QDs in hexanes under gamma-ray (γ-ray) irradiation in air[22] and another study showed that 20 keV picosecond electron pulses incident on multilayer close-packed CdSe/ZnS colloidal QD films leads to charged exciton species and multiexciton states.[23] However, the influence of high energy photons on the optical properties of colloidal QDs immobilized on a substrate remains to be explored in detail.

In this work, we report on a detailed analysis of the optical properties of sub-monolayer CdTe/CdS colloidal QDs immobilized within porous silica (PSiO$_2$) scaffolds under increasing X-ray and γ-ray exposure doses in air and nitrogen environments. Recent work demonstrated that colloidal QDs may be dispersed in a PSiO$_2$ framework with little influence on their solution phase optical properties.[24] The exceptionally high surface area of PSiO$_2$ (~200 m$^2$ cm$^{-3}$) enables the attachment of a large quantity of QDs (~10$^{14}$ QDs for 1 – 2% surface area coverage) spaced sufficiently far apart to suppress inter-QD exciton couplings,[24] which facilitates the formation of highly luminescent QD distributions that are surface-bound but still largely accessible for surface modification. In this context, using the QD-PSiO$_2$ platform, we can study QD radiation sensitivity on a solid surface that mimics a potential device configuration while preventing inter-QD exciton couplings that may otherwise encourage other avenues for non-radiative exciton annihilation and interfere with conclusions that are drawn purely from radiation induced changes in exciton dynamics.[25-27] Here, a sub-monolayer of CdTe/CdS QDs (**Supporting Information Figure S1**) are electrostatically attached within a 10 μm thick PSiO$_2$ framework (**Supporting Information Figures S2-S5**), although other types of QDs could be similarly studied using this approach. We show that cumulative 10 keV X-ray irradiation of the QD-PSiO$_2$ samples in air from 2.2 Mrad(SiO$_2$) to 6.6 Mrad(SiO$_2$) leads to an exponential decrease in QD peak emission intensity and a concurrent exponential blue-shift of the QD peak emission to higher energies due to accelerated photo-oxidative effects. Irradiating QD-PSiO$_2$ samples with higher energy and more highly penetrating 662 keV γ-rays allows for enclosure of the samples in nitrogen purged vials to suppress the effects of radiation-induced accelerated photo-oxidation; however, the QDs experience severe photodarkening with significant loss of emission intensity for less than 1 Mrad(SiO$_2$) total exposure dose. Importantly, it is shown that the effects of X-ray and γ-ray irradiation are largely reversible following a surface treatment procedure that involves the exposure of QDs to a thiol-containing solution. Consequently, this work provides a means of assessing the influence of high energy radiation and surface treatments on the photophysical properties of QDs for future QD-integrated device applications, and offers exciting avenues into developing quantifiable, low cost, flexible, large area, re-useable radiation dosimeters for space and other high-risk environments where low mass and robustness are key criteria for selecting dosimeters for extended space missions. [28-29]

## 2. Results and Discussions

*2.1 10 keV X-ray Irradiations*

**Figure 1** shows the exponential decrease in PL peak intensities for CdTe/CdS QDs attached to PSiO$_2$ scaffolds under increasing 10-keV X-ray exposure (36.7 krad(SiO$_2$)/min) in air.

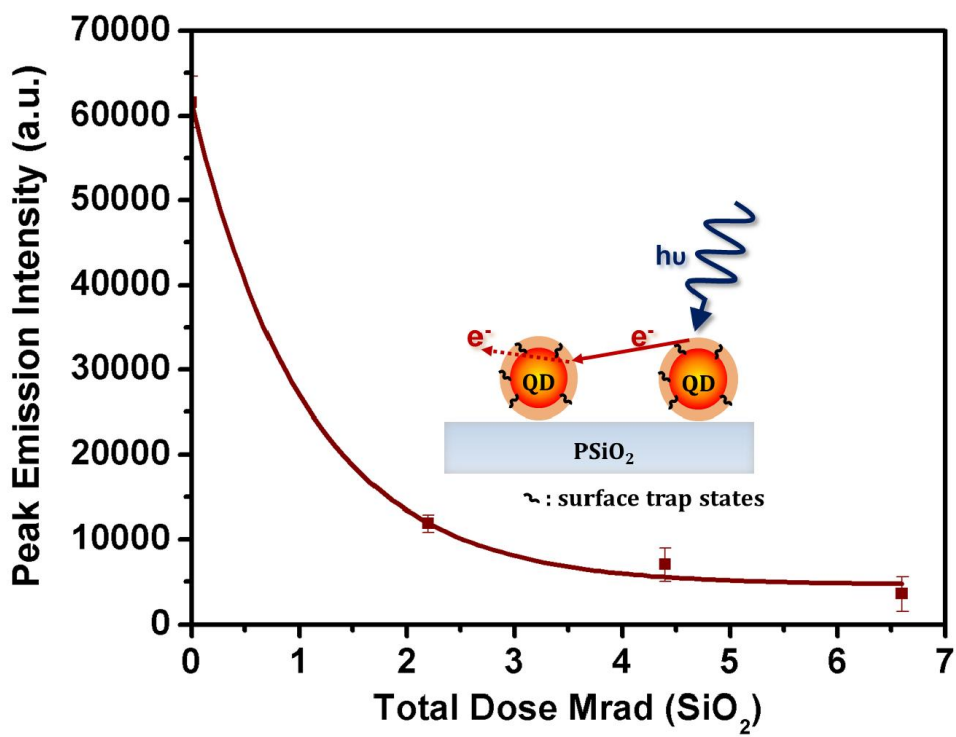

**Figure 1**. Decrease in CdTe/CdS QD PL peak intensities following increasing total exposure dose under X-ray irradiation. The solid line is indicative of a single exponential fit. Inset: Schematic representation of the interaction of 10 keV X-rays (hυ) with a QD. Ejected photoelectron is denoted as e$^-$. Radiation induced surface transformations and photocatalytic oxidation of the thiol ligands results in the formation of surface trap states.

Since the photoelectric effect dominates for low energy (<100 keV) photon interactions with materials, the interaction of 10 keV X-rays with the PSiO$_2$-QDs results in absorption of the primary photons through interactions with atoms and the generation of photoelectrons whose energy depends on the energy of the incident X-rays and the binding energies of electrons in the QDs and PSiO$_2$ framework.[30-31] As a result, the primary source of radiation induced damage for X-ray irradiated colloidal CdTe QDs is most likely through the creation of secondary electron-hole pairs formed along the track of an ejected photoelectron. As illustrated in the inset of **Figure 1**, the ejected primary photoelectron may interact with neighboring QDs resulting in the creation of secondary electron-hole pairs along the track of the primary photoelectron. Acquired charges on the QDs may initiate a permanent dark state within the QDs wherein non-radiative Auger processes are known to greatly influence the relaxation of excited electrons through Coulombic interactions with coupled holes.[32] Recent work by Zhao et. al. offers another possibility that charged QDs enter an intermediate state or "grey state" wherein they are weakly emissive with much faster radiative decays.[33] However, multiple charges present on QDs would lead to non-radiative Auger processes dominating and a complete dark state within the QDs. Additionally, in the presence of air and highly ionizing radiation, thiol-capped CdTe/CdS QDs are highly likely to undergo accelerated photooxidation of the nanocrystal/ligand complex. Previous research has demonstrated air-induced or UV-catalyzed oxidation of II-VI chalcogenides such as CdS and CdTe as well as CdSe nanocrystals coated by hydrophilic thiols. Upon photooxidation, chalcogenides such as S or Te oxidize to sulfates and oxides or sub-oxides of Te, respectively. In turn, this photooxidation results in the desorption of Cd$^{2+}$ ions or CdTe complexes from the core.[34-35] For thiol coated QDs, photogenerated holes in the QDs photocatalytically lead to oxidation of the thiol ligands and the formation of disulfides.[36] Such surface transformations or re-arrangements of surface capping agents may create trap sites present on the QD surface, core/shell interface or within core/shell structure itself and alter QD exciton dynamics through trap-mediated or Auger-assisted non-radiative carrier recombinations. Hence, the cumulative effects of increased carrier traps, photoionizations, and multiexciton creation are the most likely causes for the exponential decrease of peak QD intensities shown in **Figure 1**.

In order to investigate the role of photooxidation, X-ray photoelectron spectroscopy (XPS) measurements were carried out on an X-ray irradiated CdTe/CdS QD-PSiO$_2$ sample. As shown in **Figure 2**, the analysis focused on the S2p$^{3/2}$ and Te3d$^{5/2}$ core levels associated with CdS and CdTe bonds, respectively. After X-ray irradiation, there is an increase in the CdS binding energy from approximately 161 eV to 161.6 eV, which is consistent with oxidation, and a new peak appears at 575.9 eV, suggesting the formation of sub-oxides of Te. These results support the conclusion that the QDs are oxidized as a result of X-ray irradiation.

Previous studies have shown that the exposure of photocatalytically oxidized chalcogenide QDs to thiol-capping agents can help replace unstable disulfides that form during the photooxidation process with thiol ligands to maintain the stability of the QDs.[36] Accordingly, an X-ray irradiated CdTe/CdS QD-PSiO$_2$ sample was exposed to a solution of glutathione containing free thiols to replace the disulfides, help re-passivate surface dangling bonds of Cd ions, and possibly reduce photoionized Te atoms.[37] As shown in the XPS spectra in **Figure 2**, after the glutathione treatment, the peak at 575.9 eV is significantly decreased, suggesting removal of the sub-oxides of Te by the free thiols. In addition, two distinct peaks appear at energies near 162.1 eV and 160.9 eV. The higher energy peak at 162.1 eV might indicate the formation of sulfates while the peak near 161 eV is indicative of the re-formation of the CdS shell. These results suggest possible repassivation of the core/shell structure.

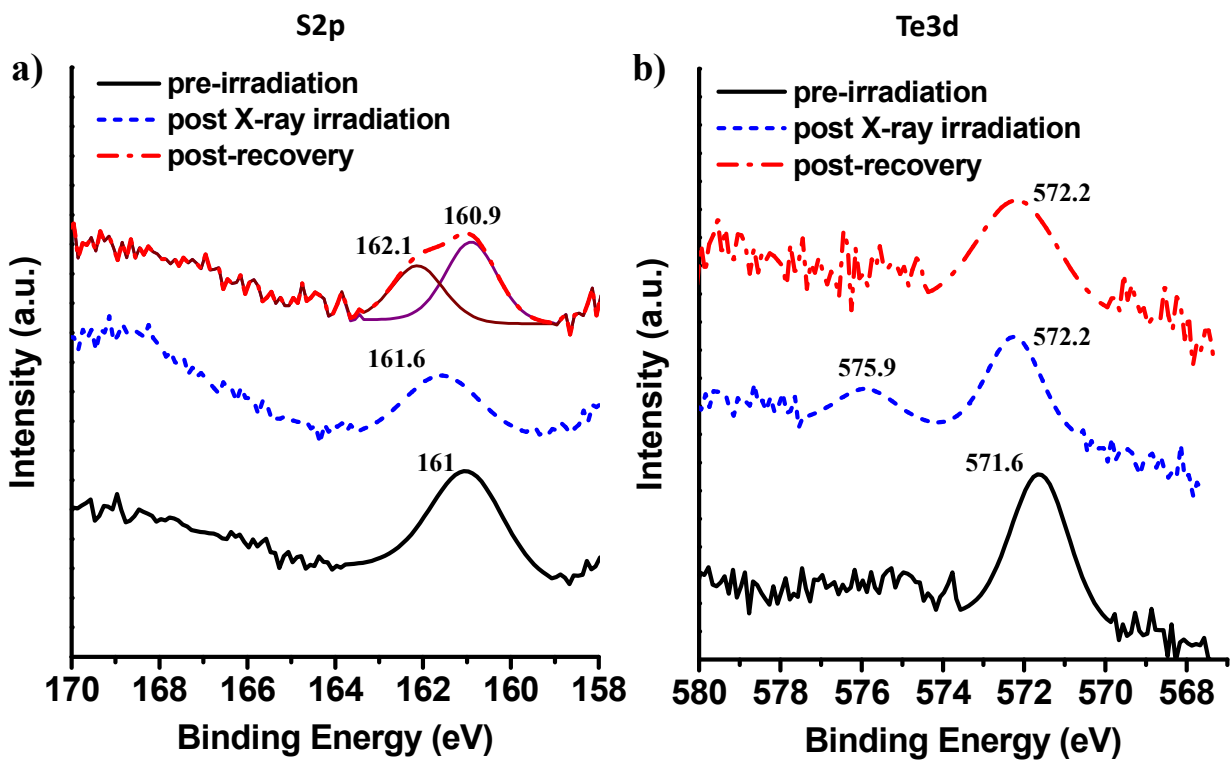

**Figure 2.** XPS spectra acquired from CdTe/CdS QDs attached to PSiO$_2$ after preparation, post X-ray irradiation, and post glutathione surface treatment: **a)** CdS S2p and **b)** Te3d$^{5/2}$ core levels. Values shown represent peak binding energies (eV) that have been calibrated to the lowest energy carbon peak at 284.8 eV.

Continuous wave photoluminescence (CWPL) measurements were then carried out to determine the effect of the glutathione treatment on the QD emission. As shown in **Figure 3,** following the surface treatment, there is an almost complete recovery of the net PL intensity, which had decreased by over an order of magnitude following a total X-ray exposure dose of 4.4 Mrad(SiO$_2$) in air. A 0.035 eV blue-shift in QD emission wavelengths is also observed following exposure to the free-thiols in solution, which may be attributed to the etching of the oxide species of S and Te formed during the X-ray irradiation. QD emission energy can be related to the effective QD diameter as given by **equation (1)** where $E_g(QD)$ is the bandgap of the QD, $h$ is Planck's constant, $a$ is the radius of the QD, $m_{eff}$ is the effective mass of an electron, and $E_g(bulk)$ is the bulk bandgap energy.[38]

$$E_g(QD) = \left(\frac{h^2}{8m_{eff}a^2}\right) + E_g(bulk) \qquad (1)$$

Given that $E_{g(bulk)}$ ~1.5 eV [39] and $m_{eff}$ ~$0.11m_e$ [40] for CdTe, we can estimate that the effective QD diameter decreases by $\Delta a$ ~130 pm for the measured $\Delta E_{g(QD)}$ ~0.035eV, given the initial QD diameter is ~3 nm. This corresponds to a decrease in effective QD diameter by about one atomic layer and increased exciton confinement that shifts the QD emission to higher energies. The magnitude of the PL peak blue-shift is related to the total ionizing dose of X-rays irradiating the QD-PSiO$_2$ samples. **Figure 4** shows the corresponding increase in QD peak emission energy after the glutathione surface treatment procedure is performed on samples exposed to X-ray irradiation at total exposure doses of 2.2 Mrad(SiO$_2$), 4.4 Mrad(SiO$_2$), and 6.6 Mrad(SiO$_2$). Control experiments show that the glutathione surface treatment procedure does not lead to increased PL intensity or a blue-shift in the PL spectrum when performed on a non-irradiated QD-PSiO$_2$ sample (**Supporting Information Figure S6)**. By correlating the degree of accelerated photo-oxidation to the X-ray total dose exposure, and re-passivating the surface, we could potentially enable a means of realizing quantitative, re-useable QD-based radiation dosimeters.

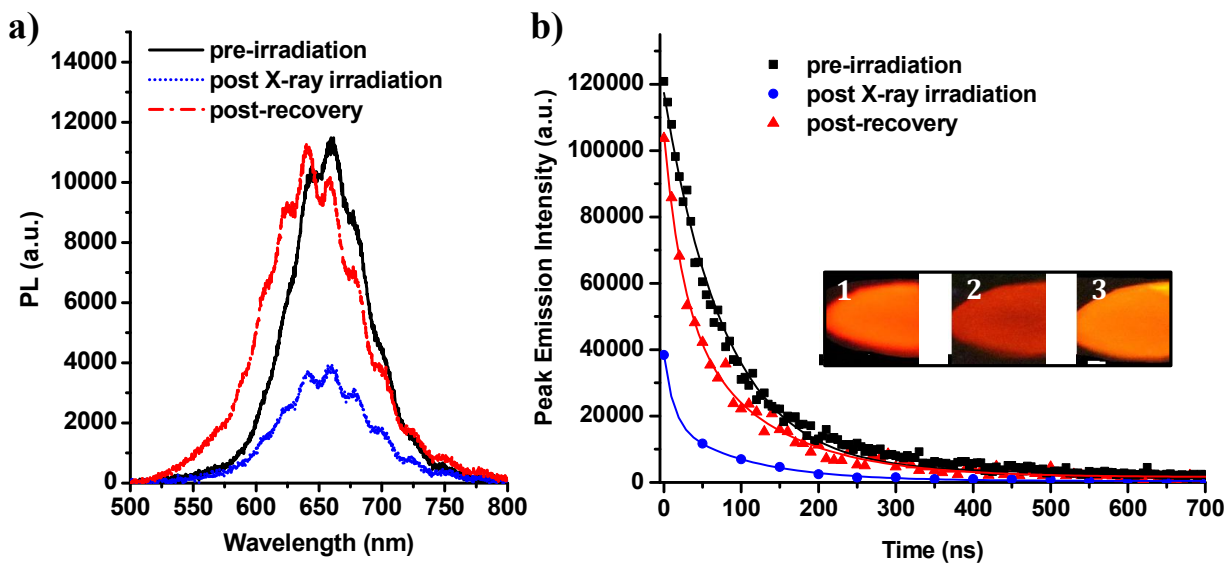

**Figure 3 a)** CWPL measurements of sub-monolayer CdTe/CdS QDs immobilized within a PSiO$_2$ thin-film as-prepared (black line), following a 4.4 Mrad(SiO$_2$) X-ray irradiation in air (blue dotted line), and post glutathione treatment (red dash-dotted line). The distinct fringes present in the spectra confirm QD infiltration and immobilization throughout the PSiO$_2$ layer.[41] **b)** TRPL measurements for CdTe/CdS QD-PSiO$_2$ samples as-prepared (black squares), following a 4.4 Mrad(SiO$_2$) X-ray irradiation (blue circles) and post glutathione treatment (red triangles). All data points are fit to a single exponential decay. Insets show camera images of samples under UV (365 nm) excitation: 1-pre-irradiation, 2-post X-ray irradiation, and 3-post glutathione treatment.

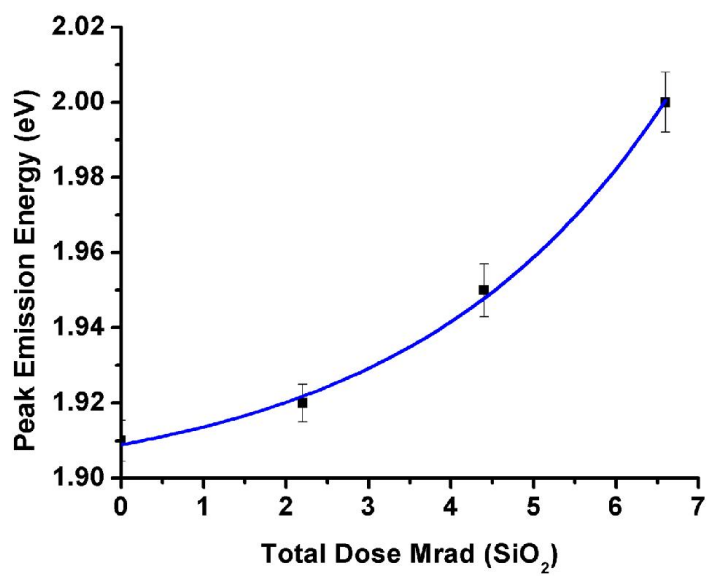

**Figure 4**. Increase in QD peak emission energy following surface recovery treatment for CdS/CdTe QD-PSiO$_2$ samples subjected to increasing X-ray doses. An exponential line fit is shown.

Time-resolved photoluminesce (TRPL) measurements were also carried out on the CdS/CdTe QD-PSiO$_2$ samples before and after X-ray irradiation and the glutathione treatment, as shown in **Figure 3b**. **Table 1** lists the average radiative lifetimes for peak QDs emission wavelengths before and after a total exposure dose of 4.4 Mrad(SiO$_2$), and following the glutathione surface treatment. The decreased lifetime measured after X-ray irradiation is consistent with the CWPL measurements (**Figure 3**) and can be attributed to a combination of increased carrier traps, photoionizations, and multiexciton creation that result from the X-ray irradiation. The increase in carrier lifetime after the glutathione treatment is likely due to a reduction in surface defect states that results from the surface treatment etching away oxide species and partly reforming the CdS shell, as suggested by the XPS (**Figure 2**) and CWPL (**Figure 3**) data.

**Table 1**[*]. QD PL lifetimes following X-ray irradiations in air

| Timeframe | PL Lifetime, τ (ns) |
|---|---|
| Initial | 78 ns ± 5 |
| Post X-ray irradiation | 47 ns ± 5 |
| Post-treatment | 66 ns ± 5 |

[*]Note: The exponential decay rates are the average values obtained from two sets of experiments.

## *2.2 662 keV γ-ray Irradiations*

Next, higher energy γ-ray irradiation (662 keV, Cesium-137 source) experiments were carried out on the QD-PSiO$_2$ samples in air and nitrogen environments to determine how the different sources of radiation and ambient conditions affect the luminescence properties of the irradiated and glutathione-treated samples. The effects of γ-irradiation on the QD-PSiO$_2$ samples in air were similar to those observed for X-ray irradiation (**Supporting Information Figure S7 and Table S1**). However, for γ-ray irradiation, a total exposure dose of only 700 krad/(SiO$_2$) was sufficient to cause near complete photodarkening of the samples accompanied by a rapid decrease in lifetime. We note that due to the lower dose rate of the γ-irradiation (0.7 krad(SiO$_2$)/min), the total exposure time of the samples to γ-rays as opposed to X-rays was significantly longer. In addition, after the glutathione surface treatment, γ-ray irradiated QD-PSiO$_2$ samples experienced a significantly higher blue-shift in peak QD emission wavelength compared to X-ray irradiated samples after the same surface treatment procedure, and the surface treatment only partially reversed the γ-radiation induced photodarkening effects (**Supporting Information Figure S7**). We attribute the blue-shift in peak QD emission to γ-ray induced photo-oxidation and the subsequent etching of oxide species during the surface treatment. The lack of complete recovery of the QD emission intensity following the surface treatment is likely due to effects resulting from partial or incomplete passivation of the surface dangling bonds after the photocatalytic oxidation of the thiolated ligands, desorption of Cd$^{2+}$ ions or CdTe complexes from the core, and permanent lattice displacement damage effects. The minimum energy for lattice displacement damage in bulk CdTe crystals is approximately 250 keV and the maximum energies of secondary electrons generated by 662 keV γ-rays through Compton scattering is approximately 480 keV.[36,37] Secondary electrons possessing energies higher than 250 keV may therefore cause displacement of atoms in the QD core/shell structure, resulting in rearrangements in the core-shell QD lattice that may lead to significant loss of QD emission intensities through the creation of several non-radiative mid-gap defect states.

Due to the deeper penetration depth of 662 keV γ-rays into materials, it is also possible to conduct γ-ray irradiation studies on QD-PSiO$_2$ samples enclosed in a nitrogen purged glass vial with a septum. This approach enables suppression of the effects of atmospheric humidity and oxygen on QD surface states that is not possible in the case of X-rays, which attenuate rapidly in glass such that any sample enclosed in a vial would be shielded from the radiation. CWPL measurements for a QD-PSiO$_2$ sample prior to and following a 700 krad/SiO$_2$ γ-irradiation in a nitrogen environment are shown in **Figure 5a**. An 80% decrease in net QD emission results from the γ-irradiation. Following exposure to the thiolated surface treatment solution, the net QD emission increases to 72% of the pre-irradiated value but is not completely reversible. Unlike the X-ray γ-irradiated samples in air, QD-PSiO$_2$ samples irradiated by γ-rays in nitrogen experience no observable blue-shift in the measured QD emission spectra following the glutathione surface treatment, confirming the absence of photo-oxidative effects on the thiolated ligands. Consequently, we can conclude that the damage to surface states is lower in the absence of accelerated photo-oxidative effects on the thiolated ligands. The recovery in CWPL observed following exposure to the free-thiol solution (**Figure 5**) hints at the possibility of reduction of the Te atoms by the free-thiols. Prior work done by Zhang et. al. also suggests an ability of hydrophilic thiols to passivate surface defect states in addition to reducing Te atoms in the case of photoionization events.[37] From this, we hypothesize that the main contributing factor towards photodarkening of the QDs under γ-irradiation in purely nitrogen environments is the creation of multi-excitons or dark states that are subsequently reversed post-treatment.

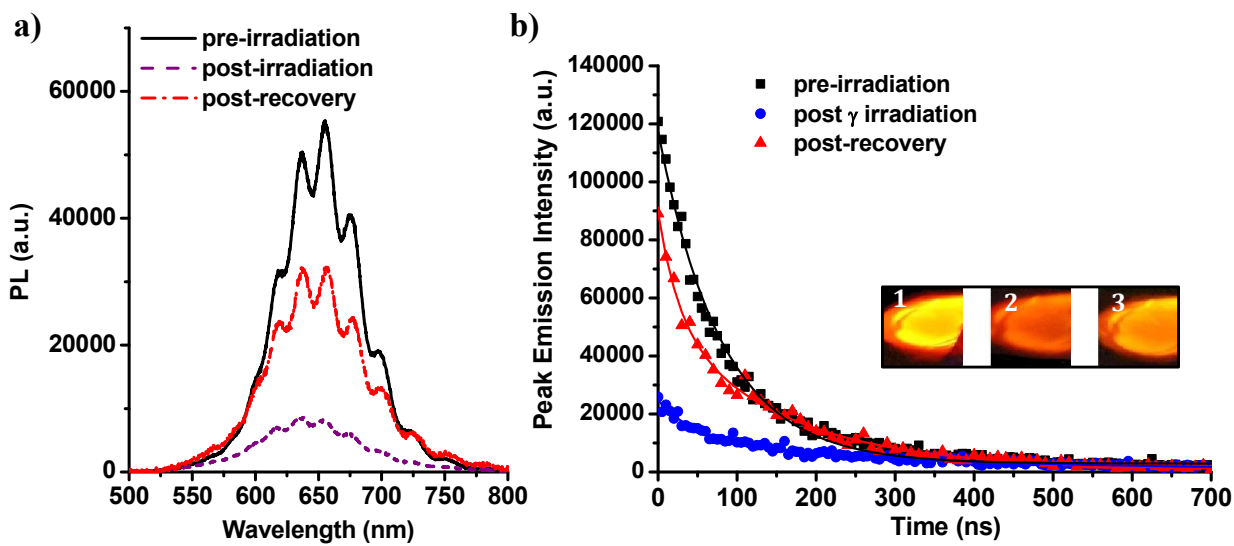

**Figure 5 a)** CWPL measurements of as-prepared sub-monolayer CdTe/CdS QDs immobilized within PSiO$_2$ thin-films (black line), following ~710 krad(SiO$_2$) γ irradiation at 0.7 krad/min(SiO$_2$) in a nitrogen environment (blue dotted line), and post glutathione treatment (red dash-dotted line). **b)** TRPL measurements of CdTe/CdS QD-PSiO$_2$ samples as-prepared (black squares), post γ irradiation at 0.7 krad/min(SiO$_2$) for 17 h in nitrogen (blue circles) and post glutathione treatment (red triangles). All data points are fit to a bi-exponential decay. Insets show camera images of samples under UV (365 nm) excitation: 1-pre-irradiation, 2-post γ-ray irradiation, and 3-post glutathione treatment.

TRPL measurements for a QD-PSiO$_2$ sample prior to and following a 700 krad/SiO$_2$ γ-irradiation in a nitrogen environment are shown in **Figure 5b**. After prolonged γ-irradiation, the exciton lifetime may be modeled with a bi-exponential fit given by $I(t) = A1 \exp(-t/\tau_1) + A2 \exp(-t/\tau_2)$, where $\tau_1$ and $\tau_2$ represent the time constants, and $A1$ and $A2$ represent the amplitudes of the components, respectively (**Table 2**).[42] The faster decay time constant $\tau_1$ is usually attributed to exciton recombinations and the much slower decay time constant $\tau_2$ may be attributed to emission of dark excitons or other trap states.[43-44] Following γ-irradiation in nitrogen, the faster PL decay component ($A1\%$) is seen to decrease accompanied by faster carrier recombination times, possibly due to increased defects. The longer PL component ($A2\%$) is seen to significantly increase accompanied by a lengthening of the PL decay time following irradiation. The reason for this is not entirely clear but could be due to a contribution from carrier trapping for extended periods of time. After the glutathione surface treatment, there is an overall decrease in both PL decay times.

**Table 2.** QD PL lifetime post γ-ray irradiations in nitrogen

| Timeframe | $A1\%$ | $\tau_1$ (ns) | $A2\%$ | $\tau_2$ (ns) |
|---|---|---|---|---|
| Initial | 79.87% | 57 ± 5 | 20.13% | 220 ± 15 |
| Post γ irradiation (N$_2$) | 46.56% | 43 ± 5 | 53.44% | 273 ± 7 |
| Post-treatment | 40.51% | 25 ± 5 | 59.49% | 151 ± 15 |

3. **Conclusions**

The effects of X-ray and γ-ray irradiation on the photophysical properties of colloidal CdTe/CdS QDs immobilized within PSiO$_2$ 3D scaffolds have been characterized. The QD emission is significantly reduced following irradiation. XPS measurements confirm photo-oxidation plays a role when irradiation is performed in air. Photoionization, carrier traps, and multiexcition generation also likely play a role in the radiation-induced photodarkening and reduced exciton lifetimes that were measured following X-ray and γ-ray irradiation of the QD-PSiO$_2$ samples. More detailed investigations of QD exciton dynamics that can resolve sub-nanosecond lifetimes would be able to shed further light on competing mechanisms of photo-oxidation, long lived carrier traps, and mid-gap defect states. Due to their higher energy, γ-rays also likely cause lattice displacements in the QDs that lead to a permanent reduction in the QD emission. CdTe/CdS QDs demonstrate near complete recovery of QD peak intensity and lifetime after X-ray irradiation when a thiol-rich surface treatment procedure is performed; partial luminescence recovery was observed for γ-irradiated samples. We believe the surface treatment is instrumental in not only reforming a CdS shell but also reducing photoionized QDs that have entered into a dark state. Quantification of the total exposure dose through monitoring changes in QD peak emission intensity and energy may enable applications in passive radiation dosimetry in high risk, high radiation environments. Lightweight and flexible QD-based thin film substrates could be realized for a variety of applications with sub-monolayer QD distributions in PSiO$_2$ 3D scaffolds that reduce carrier-trapping and charge transfer between QDs and maintain access to QDs surfaces for re-passivating solutions.

## 4. Experimental Section

*Sample Preparation:* Nanostructured porous silicon films were fabricated by electrochemical etching of boron doped p+ silicon wafers (<100>, 0.01Ω-cm, Silicon Quest) in a two-electrode configuration. A platinum wire counter-electrode and silicon wafer with an exposed area of 1.7 cm$^2$ were mounted on a silver plate in a Teflon etching cell. The electrolyte consisted of an ethanolic HF solution (3:8 v/v 49-51% aqueous HF:ethanol, Sigma Aldrich). Anodization was carried out in the dark for 334 s at an etching current density of 48 mA/cm$^2$ to form 10 μm thick nanostructured porous silicon films with average pore sizes of 25 nm. Each sample was rinsed thoroughly with ethanol and dried under a stream of nitrogen after the electrochemical etch. The samples were then cleaved in half and thermally oxidized at 1000°C in air for 3 h to form PSiO$_2$ thin-films. Little sample-to-sample variation is expected when analyzing the effects of radiation on irradiated versus control sample halves that originated from the same PSiO$_2$ film.

The PSiO$_2$ samples were incubated for 10 minutes in 3% poly(diallyldimethylammonium chloride) (PDDA) aqueous solution at pH = 3.0, followed by a deionized (DI) water rinse to remove excess molecules. PDDA molecules (~1 nm) impart a positive charge to the PSiO$_2$ substrates upon attachment.

CdTe/CdS QD preparation method: CdTe QDs were synthesized according to a modified procedure.[45] Briefly, 40 mg of NaBH$_4$ were dissolved in 2 mL of H$_2$O contained in a small vial with a septum and cooled with ice. 64 mg of Te powder were added to a solution of NaBH$_4$ and stirred until completely dissolved under constant cooling conditions and slow Ar flow. The resulting clear solution of NaHTe was transferred into 50 mL of degassed water. At the same time, 46 mg of cadmium chloride and 122 mg of glutathione (GSH) were dissolved in 50 mL of H$_2$O. The pH of the resulting Cd-GSH complex was adjusted to 10 by adding 1M NaOH solution dropwise. All solutions were purged for about 30 min with Ar before further use. CdTe QDs were prepared by injection of 5 mL NaHTe solution into a Cd-GSH solution, which was then heated at 100 °C for 1 hour. Additional degree of QDs surface passivation was achieved by overcoating synthesized CdTe QDs with CdS.[46-47] Briefly, 3.8 mg of thioacetamide (TAA) were added to synthesized CdTe QDs at 100 °C and the mixture was refluxed for 40 min. After the synthesis, the QDs solution was cooled down to room temperature.

Negatively charged CdTe/CdS QDs electrostatically bind to the positively charged PDDA coated PSiO$_2$ surface during a 20 minute incubation period. Unattached CdTe/CdS QDs were then washed away with thorough rinsing under DI water.

*Optical Characterization:* Absorbance and reflectance spectra were measured at room temperature with a Varian Cary 5000 UV-VIS-NIR spectrophotometer at a step size of 0.5 nm. Absorbance spectra were collected over a wavelength range of 300 nm – 800 nm. Reflectance spectra were collected over a wavelength range of 500 nm – 2000 nm using a spot size of ~6 mm. CWPL measurements were made using an Ar-Kr laser (Coherent Innova 70C) operating at a wavelength of 488 nm and power of 3 mW as the excitation source and a CCD spectrometer (Ocean Optics USB4000) fitted with a 1000 μm diameter optical fiber to record visible QD emission from the samples between 500 nm and 800 nm. TRPL measurements were carried out with an intensified CCD detector (iDUS490A, Andor Technology) attached to a spectrograph (Shamrock, SR303i, Andor Technology). A Nd:YAG Q-switched laser (Minilite-10, Continuum Inc.) operating at a wavelength of 355 nm in low power mode (10 mW), with 10 ns pulse duration and 10 Hz repetition rate was used as the excitation source for the TRPL experiments.

*X-ray Irradiation:* PSi-QD samples were irradiated with 10 keV X-rays at a dose rate of 36.7 krad/min($SiO_2$) in an ARACOR 4100 for exposure times varying from 1 h to 3 h in ambient environments.

*Gamma-ray Irradiation:* A Cesium-137 source was used for 662 keV γ-irradiation of QD-$PSiO_2$ samples at a dose rate of 0.7 krad/min($SiO_2$) for a total dose of ~700 krad($SiO_2$). For some experiments, the samples were sealed in nitrogen purged glass vials prior to being irradiated to minimize the effects of oxygen and moisture on QD exciton dynamics.

*Surface Re-Passivation:* 100 μL of a freshly prepared aqueous glutathione solution (0.3 mM, pH = 7.3) was pipetted onto irradiated QD samples for varying amounts of time (5 min, 10 min, 20 min, 30 min, and 40 min). The samples were then rinsed with DI water and dried under nitrogen. An incubation time of 25 min was found to be sufficient to achieve almost complete recovery of QD emission intensities.

**Acknowledgements**

This work was funded in part by the Defense Threat Reduction Agency (Grant no. HDTRA1-10-1-0041). The authors thank the Vanderbilt Institute for Nanoscale Science and Engineering for equipment usage and facilities that were renovated under NSF ARI-R2 DMR-0963361. We thank J. E. Macdonald for technical insights into the photophysical properties of quantum dots and B. Schmidt for helpful discussions related to XPS. The authors gratefully acknowledge M. McCurdy for operating the irradiators, S. Avanesyan and R. F. Haglund for assistance with and access to the pulsed laser setup, and S. M. Harrell for useful technical discussions.

**Supporting Information**

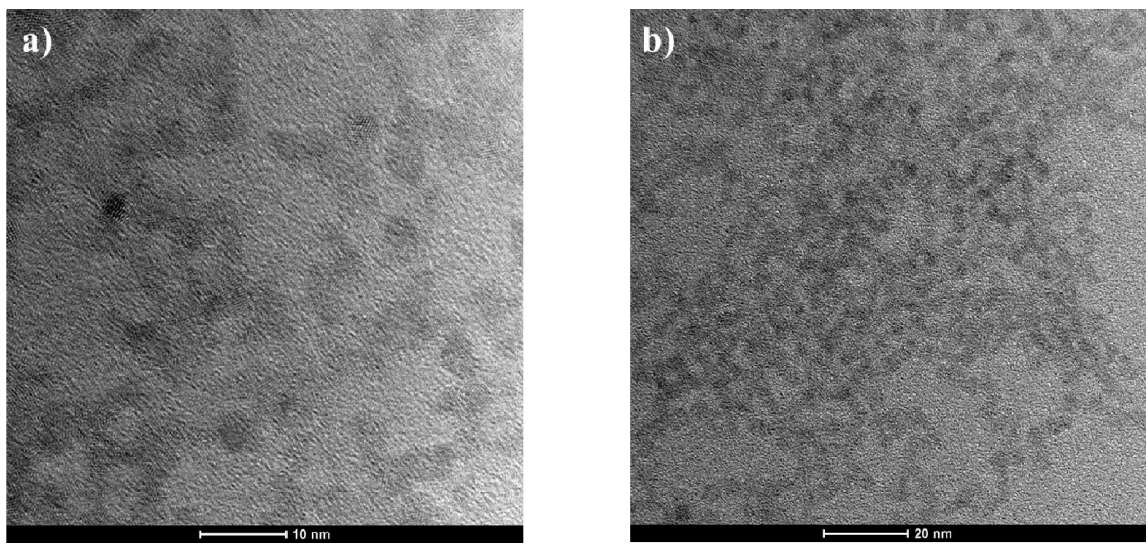

**Figure S1**. TEM images of the ~3.5 nm CdTe/CdS QDs.

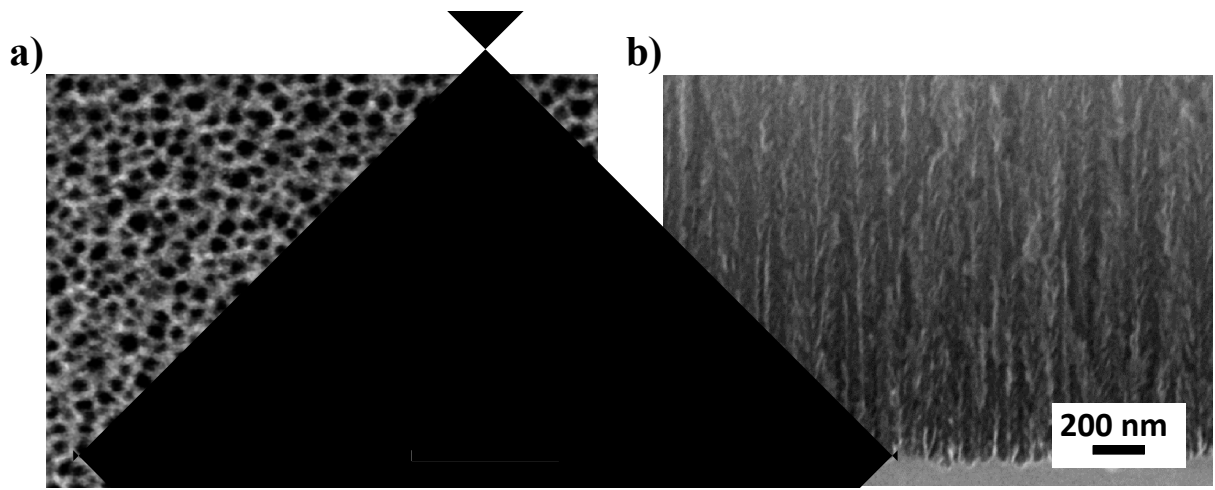

**Figure S2.** SEM images of freshly prepared PSiO$_2$ showing a) top view and b) cross-sectional profile. The average pore sizes are ~25 nm with inter-pore nano-wall dimensions of ~9 nm on average. Nanowire branches present along the pore lengths are < 5 nm on average.

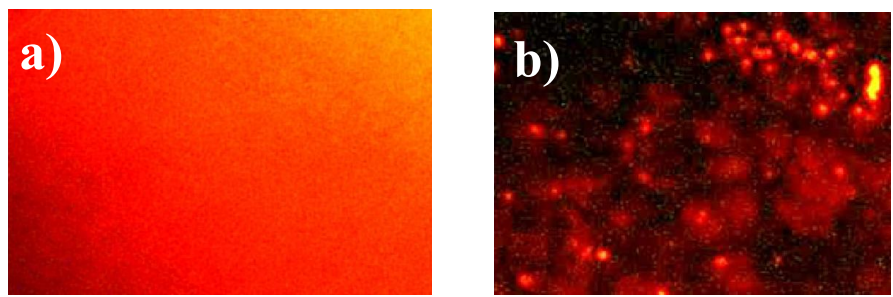

**Figure S3.** Fluorescence microscopy images of CdTe/CdS QDs attached to PDDA coated substrates: a) 10 μm thick PSiO$_2$ film and b) flat Si sample. Due to the large internal surface area of the PSiO$_2$ film, a significantly larger quantity of QDs is attached to PSiO$_2$ compared to flat Si, as is indicated by the brighter fluorescence microscopy image.

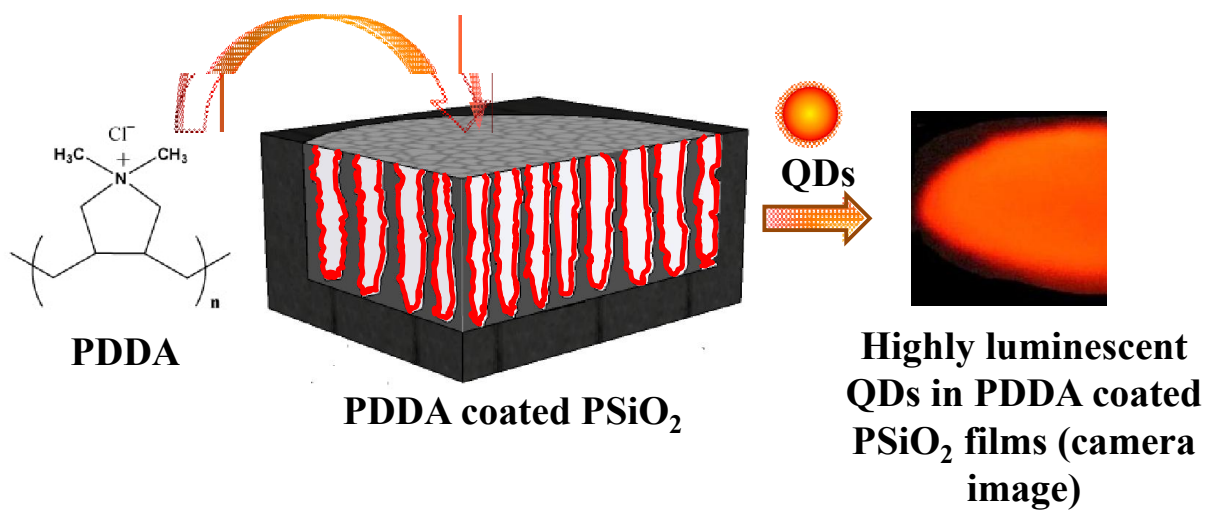

**Figure S4.** Schematic illustration of the attachment of CdTe/CdS QDs to PDDA coated PSiO$_2$ film and a camera image of the sample under UV lamp excitation at 365 nm.

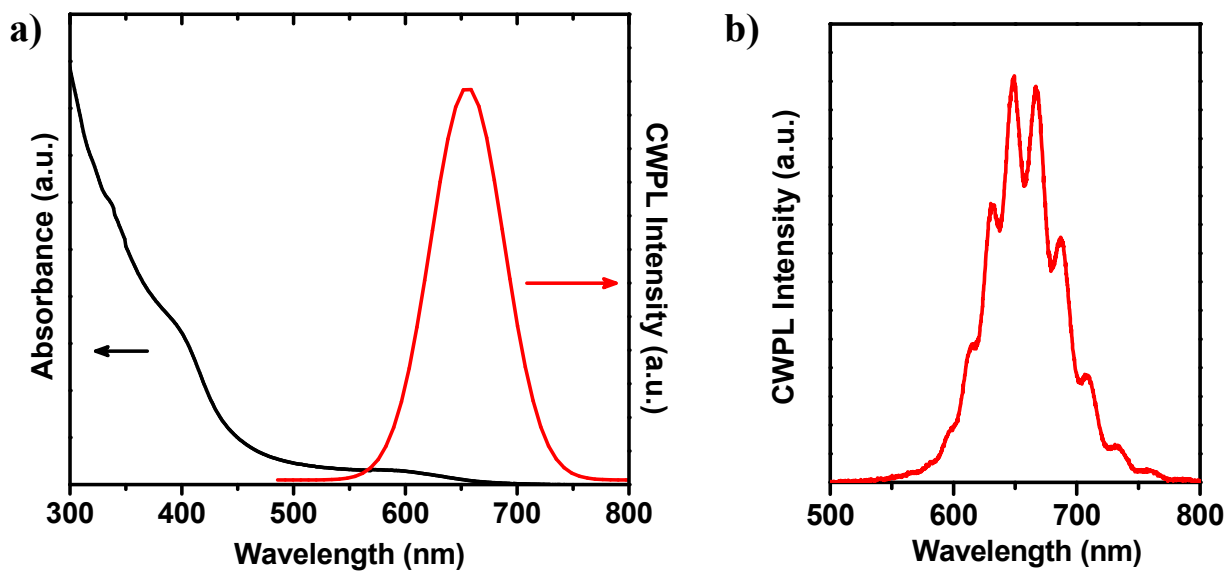

**Figure S5. a**) Absorbance and CWPL spectrum for CdTe/CdS QDs. **b**) CWPL spectrum of sub-monolayer CdTe/CdS QDs immobilized in a PSiO$_2$ thin-film.

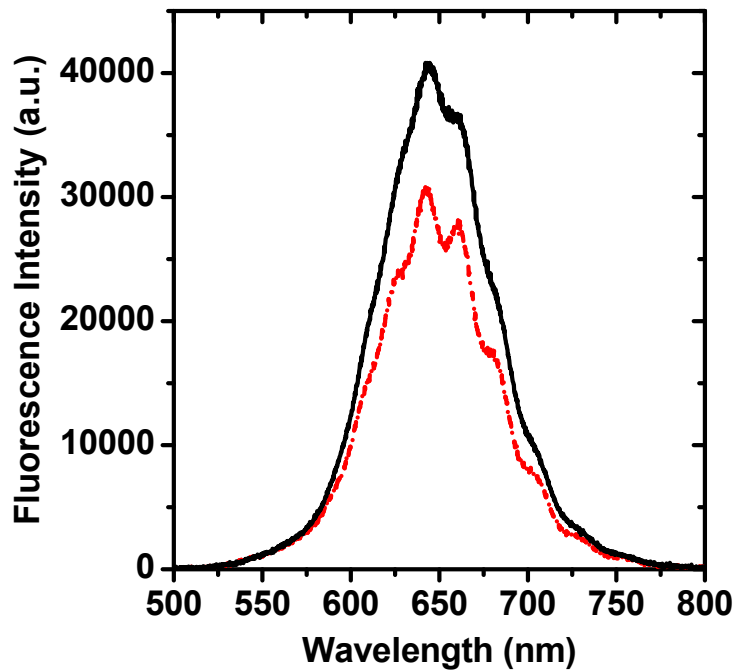

**Figure S6.** Fluorescence spectra of a QD-PSiO$_2$ sample before (black solid line) and after (red dashed line) incubation in 0.3 mM aqueous glutathione solution for 20 min. The latter measurement was performed after a 2 h time interval that corresponds to the time duration of a 4.04 Mrad(SiO$_2$) total ionizing dose X-ray irradiation carried out at a dose rate of 36.7 krad/min(SiO$_2$).. The QD-PSiO$_2$ samples were freshly prepared prior to performing the experiments.

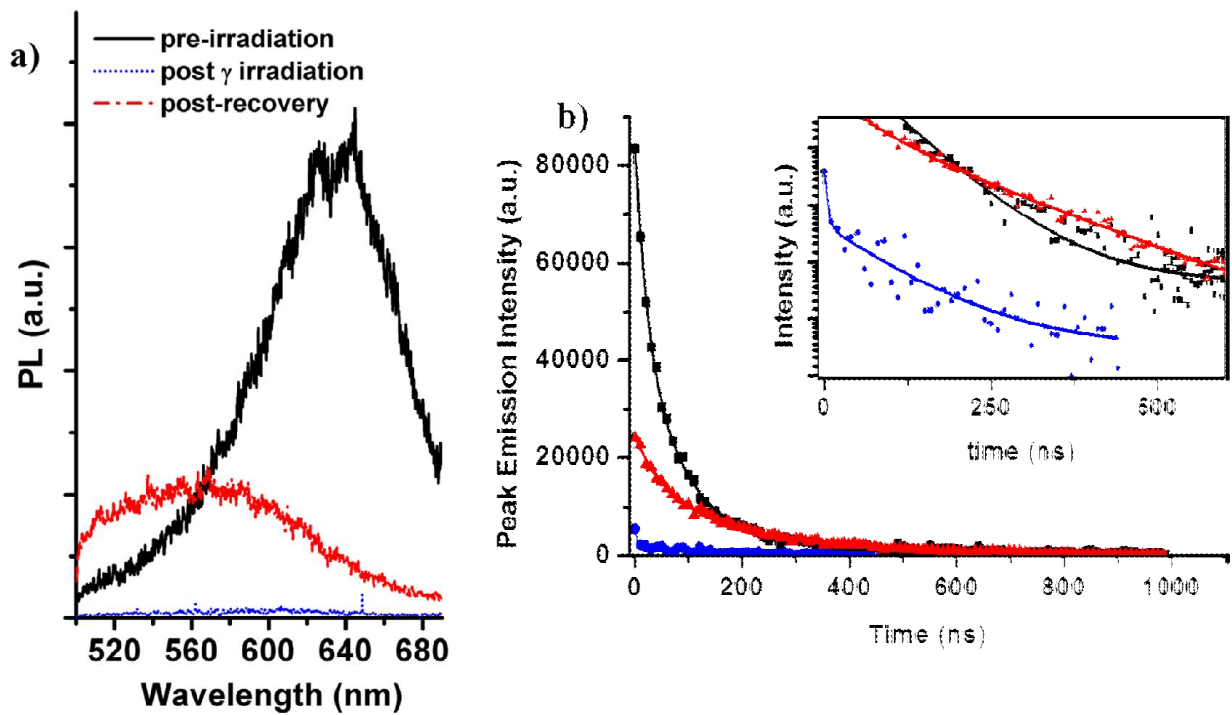

**Figure S7. a)** CWPL measurements of as-prepared sub-monolayer CdTe/CdS QDs immobilized within PSiO$_2$ thin-films (black line), post ~700 krad(SiO$_2$) γ-irradiation at 0.7 krad/min(SiO$_2$) in air (blue dotted line), and post surface treatment (red dash-dotted line). **b)** TRPL measurements of CdTe/CdS QD-PSiO$_2$ samples as-prepared (black squares), post γ-irradiation at 0.7 krad/min(SiO$_2$) for 17 h in nitrogen (blue circles) and post surface treatment (red triangles). All data points are fit to a bi-exponential decay. The inset shows a zoomed-in plot for the QD carrier lifetime curves. Post-γ irradiation, the bi-exponential fit is clearly evident.

**Table S1.** QD exciton lifetimes post γ-ray irradiations in air.

| | |
|---|---|
| Initial | ~80 ns |
| Post γ irradiation (Air) | ~4 ns + 120 ns |
| Post-treatment | ~53 ns + 216 ns |

1.	Lee, Y.-L.; Lo, Y.-S., Highly Efficient Quantum-Dot-Sensitized Solar Cell Based on Co-Sensitization of CdS/CdSe. *Advanced Functional Materials* **2009,** *19* (4), 604-609.
2.	Schaller, R. D.; Sykora, M.; Pietryga, J. M.; Klimov, V. I., Seven Excitons at a Cost of One: Redefining the Limits for Conversion Efficiency of Photons into Charge Carriers. *Nano Lett* **2006,** *6* (3), 424-429.
3.	Gao, J.; Luther, J. M.; Semonin, O. E.; Ellingson, R. J.; Nozik, A. J.; Beard, M. C., Quantum Dot Size Dependent J−V Characteristics in Heterojunction ZnO/PbS Quantum Dot Solar Cells. *Nano Lett* **2011,** *11* (3), 1002-1008.
4.	Lu, S.; Lingley, Z.; Asano, T.; Harris, D.; Barwicz, T.; Guha, S.; Madhukar, A., Photocurrent Induced by Nonradiative Energy Transfer from Nanocrystal Quantum Dots to Adjacent Silicon Nanowire Conducting Channels: Toward a New Solar Cell Paradigm. *Nano Lett* **2009,** *9* (12), 4548-4552.
5.	Gao, J.; Perkins, C. L.; Luther, J. M.; Hanna, M. C.; Chen, H.-Y.; Semonin, O. E.; Nozik, A. J.; Ellingson, R. J.; Beard, M. C., n-Type Transition Metal Oxide as a Hole Extraction Layer in PbS Quantum Dot Solar Cells. *Nano Lett* **2011,** *11* (8), 3263-3266.
6.	McDonald, S. A.; Konstantatos, G.; Zhang, S.; Cyr, P. W.; Klem, E. J. D.; Levina, L.; Sargent, E. H., Solution-processed PbS quantum dot infrared photodetectors and photovoltaics. *Nat Mater* **2005,** *4* (2), 138-142.
7.	Sukhovatkin, V.; Hinds, S.; Brzozowski, L.; Sargent, E. H., Colloidal Quantum-Dot Photodetectors Exploiting Multiexciton Generation. *Science* **2009,** *324* (5934), 1542-1544.
8.	Gao, X.; Cui, Y.; Levenson, R. M.; Chung, L. W.; Nie, S., In vivo cancer targeting and imaging with semiconductor quantum dots. *Nature biotechnology* **2004,** *22* (8), 969-976.
9.	Rizvi, S. B.; Ghaderi, S.; Keshtgar, M.; Seifalian, A. M., Semiconductor quantum dots as fluorescent probes for in vitro and in vivo bio-molecular and cellular imaging. *Nano Reviews* **2010,** *1*, 10.3402/nano.v3401i3400.5161.
10.	Yang, J.; Dave, S. R.; Gao, X., Quantum Dot Nanobarcodes: Epitaxial Assembly of Nanoparticle−Polymer Complexes in Homogeneous Solution. *Journal of the American Chemical Society* **2008,** *130* (15), 5286-5292.
11.	Zhao, J.; Bardecker, J. A.; Munro, A. M.; Liu, M. S.; Niu, Y.; Ding, I. K.; Luo, J.; Chen, B.; Jen, A. K. Y.; Ginger, D. S., Efficient CdSe/CdS Quantum Dot Light-Emitting Diodes Using a Thermally Polymerized Hole Transport Layer. *Nano Lett* **2006,** *6* (3), 463-467.
12.	Sun, Q.; Wang, Y. A.; Li, L. S.; Wang, D.; Zhu, T.; Xu, J.; Yang, C.; Li, Y., Bright, multicoloured light-emitting diodes based on quantum dots. *Nat Photon* **2007,** *1* (12), 717-722.
13.	Caruge, J. M.; Halpert, J. E.; Wood, V.; Bulovic, V.; Bawendi, M. G., Colloidal quantum-dot light-emitting diodes with metal-oxide charge transport layers. *Nat Photon* **2008,** *2* (4), 247-250.
14.	Cho, K.-S.; Lee, E. K.; Joo, W.-J.; Jang, E.; Kim, T.-H.; Lee, S. J.; Kwon, S.-J.; Han, J. Y.; Kim, B.-K.; Choi, B. L.; Kim, J. M., High-performance crosslinked colloidal quantum-dot light-emitting diodes. *Nat Photon* **2009,** *3* (6), 341-345.
15.	Shirasaki, Y.; Supran, G. J.; Bawendi, M. G.; Bulovic, V., Emergence of colloidal quantum-dot light-emitting technologies. *Nat Photon* **2013,** *7* (1), 13-23.
16.	Campbell, I. H.; Crone, B. K., Quantum-Dot/Organic Semiconductor Composites for Radiation Detection. *Advanced Materials* **2006,** *18* (1), 77-79.
17.	Létant, S. E.; Wang, T. F., Semiconductor Quantum Dot Scintillation under γ-Ray Irradiation. *Nano Lett* **2006,** *6* (12), 2877-2880.
18.	Stroh, M.; Zimmer, J. P.; Duda, D. G.; Levchenko, T. S.; Cohen, K. S.; Brown, E. B.; Scadden, D. T.; Torchilin, V. P.; Bawendi, M. G.; Fukumura, D., Quantum dots spectrally distinguish multiple species within the tumor milieu in vivo. *Nature medicine* **2005,** *11* (6), 678-682.
19.	Kang, Z.; Zhang, Y.; Menkara, H.; Wagner, B. K.; Summers, C. J.; Lawrence, W.; Nagarkar, V., CdTe quantum dots and polymer nanocomposites for x-ray scintillation and imaging. *Applied Physics Letters* **2011,** *98* (18), 181914-181914-181913.

20. Guffarth, F.; Heitz, R.; Geller, M.; Kapteyn, C.; Born, H.; Sellin, R.; Hoffmann, A.; Bimberg, D.; Sobolev, N. A.; Carmo, M. C., Radiation hardness of InGaAs/GaAs quantum dots. *Applied Physics Letters* **2003,** *82* (12), 1941-1943.
21. Leon, R.; Swift, G. M.; Magness, B.; Taylor, W. A.; Tang, Y. S.; Wang, K. L.; Dowd, P.; Zhang, Y. H., Changes in luminescence emission induced by proton irradiation: InGaAs/GaAs quantum wells and quantum dots. *Applied Physics Letters* **2000,** *76* (15), 2074-2076.
22. Withers, N. J.; Sankar, K.; Akins, B. A.; Memon, T. A.; Gu, T.; Gu, J.; Smolyakov, G. A.; Greenberg, M. R.; Boyle, T. J.; Osiński, M., Rapid degradation of CdSe ∕ ZnS colloidal quantum dots exposed to gamma irradiation. *Applied Physics Letters* **2008,** *93* (17), 173101.
23. Padilha, L. A.; Bae, W. K.; Klimov, V. I.; Pietryga, J. M.; Schaller, R. D., Response of Semiconductor Nanocrystals to Extremely Energetic Excitation. *Nano Lett* **2013,** *13* (3), 925-932.
24. Gaur, G.; Koktysh, D. S.; Fleetwood, D. M.; Weller, R. A.; Reed, R. A.; Weiss, S. M., Influence of interfacial oxide on the optical properties of single layer CdTe/CdS quantum dots in porous silicon scaffolds. *Applied Physics Letters* **2015,** *107* (6), 063106.
25. Orlova, A.; Gromova, Y. A.; Maslov, V.; Andreeva, O.; Baranov, A.; Fedorov, A.; Prudnikau, A.; Artemyev, M.; Berwick, K., Reversible photoluminescence quenching of CdSe/ZnS quantum dots embedded in porous glass by ammonia vapor. *Nanotechnology* **2013,** *24* (33), 335701.
26. Crisp, R. W.; Schrauben, J. N.; Beard, M. C.; Luther, J. M.; Johnson, J. C., Coherent Exciton Delocalization in Strongly Coupled Quantum Dot Arrays. *Nano Lett* **2013,** *13* (10), 4862-4869.
27. Sun, L.; Choi, J. J.; Stachnik, D.; Bartnik, A. C.; Hyun, B.-R.; Malliaras, G. G.; Hanrath, T.; Wise, F. W., Bright infrared quantum-dot light-emitting diodes through inter-dot spacing control. *Nat Nano* **2012,** *7* (6), 369-373.
28. Tawara, H.; Masukawa, M.; Nagamatsu, A.; Kitajo, K.; Kumagai, H.; Yasuda, N., Characteristics of Mg2SiO4:Tb (TLD-MSO-S) relevant for space radiation dosimetry. *Radiation Measurements* **2011,** *46* (8), 709-716.
29. Nagamatsu, A.; Murakami, K.; Kitajo, K.; Shimada, K.; Kumagai, H.; Tawara, H., Area radiation monitoring on ISS Increments 17 to 22 using PADLES in the Japanese Experiment Module Kibo. *Radiation Measurements* **2013,** *59* (0), 84-93.
30. Attix, F. H.; Roesch, W. C.; Tochilin, E., *Radiation Dosimetry: Fundamentals*. Academic Press: 1968.
31. Gaur, G.; Koktysh, D.; Fleetwood, D. M.; Reed, R. A.; Weller, R. A.; Weiss, S. M. In *Effects of x-ray and gamma-ray irradiation on the optical properties of quantum dots immobilized in porous silicon*, 2013; pp 87252D-87252D-87258.
32. Gao, Y.; Sandeep, C. S. S.; Schins, J. M.; Houtepen, A. J.; Siebbeles, L. D. A., Disorder strongly enhances Auger recombination in conductive quantum-dot solids. *Nat Commun* **2013,** *4*.
33. Zhao, J.; Nair, G.; Fisher, B. R.; Bawendi, M. G., Challenge to the Charging Model of Semiconductor-Nanocrystal Fluorescence Intermittency from Off-State Quantum Yields and Multiexciton Blinking. *Physical Review Letters* **2010,** *104* (15), 157403.
34. Alivisatos, A. P., Perspectives on the Physical Chemistry of Semiconductor Nanocrystals. *The Journal of Physical Chemistry* **1996,** *100* (31), 13226-13239.
35. Derfus, A. M.; Chan, W. C. W.; Bhatia, S. N., Probing the Cytotoxicity of Semiconductor Quantum Dots. *Nano Lett* **2004,** *4* (1), 11-18.
36. Aldana, J.; Wang, Y. A.; Peng, X., Photochemical Instability of CdSe Nanocrystals Coated by Hydrophilic Thiols. *Journal of the American Chemical Society* **2001,** *123* (36), 8844-8850.
37. Zhang, Y.; Li, Y.; Yan, X.-P., Photoactivated CdTe/CdSe Quantum Dots as a Near Infrared Fluorescent Probe for Detecting Biothiols in Biological Fluids. *Anal Chem* **2009,** *81* (12), 5001-5007.
38. Nirmal, M.; Brus, L., Luminescence photophysics in semiconductor nanocrystals. *Accounts of Chemical Research* **1999,** *32* (5), 407-414.
39. Fonthal, G.; Tirado-Mejía, L.; Marín-Hurtado, J. I.; Ariza-Calderón, H.; Mendoza-Alvarez, J. G., Temperature dependence of the band gap energy of crystalline CdTe. *Journal of Physics and Chemistry of Solids* **2000,** *61* (4), 579-583.

40. Marple, D. T. F., Effective Electron Mass in CdTe. *Physical Review* **1963,** *129* (6), 2466-2470.
41. Gaur, G.; Koktysh, D. S.; Weiss, S. M., Immobilization of Quantum Dots in Nanostructured Porous Silicon Films: Characterizations and Signal Amplification for Dual-Mode Optical Biosensing. *Advanced Functional Materials* **2013**.
42. Zeng, Q.; Kong, X.; Sun, Y.; Zhang, Y.; Tu, L.; Zhao, J.; Zhang, H., Synthesis and Optical Properties of Type II CdTe/CdS Core/Shell Quantum Dots in Aqueous Solution via Successive Ion Layer Adsorption and Reaction. *The Journal of Physical Chemistry C* **2008,** *112* (23), 8587-8593.
43. Efros, A. L.; Rosen, M.; Kuno, M.; Nirmal, M.; Norris, D. J.; Bawendi, M., Band-edge exciton in quantum dots of semiconductors with a degenerate valence band: Dark and bright exciton states. *Phys Rev B* **1996,** *54* (7), 4843-4856.
44. Wang, X.; Qu, L.; Zhang, J.; Peng, X.; Xiao, M., Surface-Related Emission in Highly Luminescent CdSe Quantum Dots. *Nano Lett* **2003,** *3* (8), 1103-1106.
45. Liu, Y. F.; Yu, J. S., Selective synthesis of CdTe and high luminescence CdTe/CdS quantum dots: The effect of ligands. *J. Colloid Interface Sci.* **2009,** *333* (2), 690-698.
46. Wang, Y.; Tang, Z. Y.; Correa-Duarte, M. A.; Pastoriza-Santos, I.; Giersig, M.; Kotov, N. A.; Liz-Marzan, L. M., Mechanism of strong luminescence photoactivation of citrate-stabilized water-soluble nanoparticles with CdSe cores. *J. Phys. Chem. B* **2004,** *108* (40), 15461-15469.
47. Qian, H.; Dong, C.; Peng, J.; Qiu, X.; Xu, Y.; Ren, J., High-quality and water-soluble near-infrared photolumineseent CdHgTe/CdS quantum dots prepared by adjusting size and composition. *J. Phys. Chem. C* **2007,** *111* (45), 16852-16857.